\documentclass[12pt]{article}
\usepackage{amssymb}
\usepackage{amsmath}
\usepackage{ntheorem}
\usepackage{graphicx}
\usepackage{multirow}
\usepackage{float}
\usepackage{authblk}
\usepackage[font=small,labelfont=bf,tableposition=top]{caption}
\usepackage{threeparttable}
\usepackage[ruled, linesnumbered]{algorithm2e}
\def\bfm#1{\mbox{\boldmath$#1$}}
\numberwithin{equation}{section}

\title{A New Projection Pursuit Index for Big Data}
\author[1]{Yajie Duan}
\author[1]{Javier Cabrera}
\author[2]{Birol Emir}
\affil[1]{Department of Statistics, Rutgers University, NJ, USA}
\affil[2]{Pfizer Inc, New York, NY, USA}

\begin{document}
\maketitle
\voffset -0.5truecm \hoffset -1.5truecm


\renewcommand{\theequation}{\thesection.\arabic {equation}}

\newtheorem{corollary}{Corollary}
\newtheorem{definition}{Definition}
\newtheorem{example}{Example}
\newtheorem{lemma}{Lemma}
\newtheorem{property}{Property}
\newtheorem{remark}{Remark}
\newtheorem{theorem}{Theorem}
\newtheorem{Table}{Table}


\newcommand{\bbB}{{\mathbb{B}}}
\newcommand{\bbC}{{\mathbb{C}}}
\newcommand{\bbD}{{\mathbb{D}}}
\newcommand{\bbE}{{\mathbb{E}}}
\newcommand{\bbI}{{\mathbb{I}}}
\newcommand{\bbJ}{{\mathbb{J}}}
\newcommand{\bbK}{{\mathbb{K}}}
\newcommand{\bbN}{{\mathbb{N}}}
\newcommand{\bbO}{{\mathbb{O}}}
\newcommand{\bbR}{{\mathbb{R}}}
\newcommand{\bbS}{{\mathbb{S}}}
\newcommand{\bbT}{{\mathbb{T}}}
\newcommand{\bbV}{{\mathbb{V}}}
\newcommand{\bbX}{{\mathbb{X}}}
\newcommand{\bbY}{{\mathbb{Y}}}


\newcommand{\cA}{{\cal A}}         \newcommand{\dA}[1]{{\cal A}_{#1}}
\newcommand{\cB}{{\cal B}}         \newcommand{\dB}[1]{{\cal B}_{#1}}
\newcommand{\cI}{{\cal I}}
\newcommand{\cJ}{{\cal J}}
\newcommand{\cK}{{\cal K}}
\newcommand{\cN}{{\cal N}}
\newcommand{\cO}{{\cal O}}
\newcommand{\cS}{{\cal S}}         \newcommand{\dS}[1]{{\cal S}_{#1}}
\newcommand{\cR}{{\cal R}}
\newcommand{\cT}{{\cal T}}
\newcommand{\cU}{{\cal U}}
\newcommand{\cV}{{\cal V}}
\newcommand{\cX}{{\cal X}}
\newcommand{\cY}{{\cal Y}}
\newcommand{\cZ}{{\cal Z}}


\newcommand{\f}[1]{f_{_{\tiny\rm #1}}}  \newcommand{\hf}[1]{\hat{f}_{#1}}
\newcommand{\tf}[2]{f_{{#1}}^{{#2}}}

\newcommand{\h}[1]{h_{_{\tiny\rm #1}}}
\newcommand{\n}[1]{n_{_{\tiny\rm #1}}}
\newcommand{\q}[2]{q_{_{\tiny\rm #1\!, \, #2}}}
\newcommand{\del}[1]{\delta_{_{\tiny\rm #1}}}
\newcommand{\pii}[2]{\pi_{_{\tiny\rm #1\!, \, #2}}}

\newcommand{\y}[2]{y_{#1}^{\rm #2}}
\newcommand{\yb}[1]{{\bar{y}^{\rm #1}}}
\newcommand{\Y}[1]{Y_{\rm #1}}

\newcommand{\YY}[1]{Y_{\! \mbox{\scriptsize #1}}}

\newcommand{\no}[2]{\|#1\|_{_{#2}}}
\newcommand{\nor}[3]{\|#1\|_{_{#2}}^{#3}}

\newcommand{\bismall}[2]{(\!\!\begin{array}{c} #1 \\ \vspace*{-0.6cm} \\ #2 \end{array} \!\! )}
\newcommand{\bi}[2]{\Big(\!\!\begin{array}{c} #1 \\ #2 \end{array} \!\!\Big)}
\newcommand{\twoc}[2]{\bigg(\!\!\begin{array}{c} #1 \\ #2  \end{array} \!\!\bigg)}
\newcommand{\twolr}[2]{\left(\!\!\begin{array}{c} #1 \\ #2  \end{array} \!\!\right)}

\newcommand{\twob}[2]{\lbrace {#1 \atop #2} \rbrace}
\newcommand{\twoB}[2]{\bigg\{\!\!\begin{array}{c} #1 \\ #2  \end{array} \!\!\bigg\}}

\newcommand{\fourc}[4]{\bigg(\!\!\begin{array}{cc} #1 & #2 \\ #3 & #4 \end{array} \!\!\bigg)}
\newcommand{\fourlr}[4]{\left(\!\!\begin{array}{cc} #1 & #2 \\ #3 & #4 \end{array} \!\!\right)}

\newcommand{\threec}[3]{\left(\!\!\begin{array}{c} #1 \\ #2 \\ #3  \end{array} \!\!\right)}
\newcommand{\threetwo}[6]{\left(\!\!\begin{array}{cc}
#1 & #2 \\
#3 & #4 \\
#5 & #6
\end{array} \!\!\right)}
\newcommand{\threethree}[9]{\left(\!\!\begin{array}{ccc}
#1 & #2 & #3 \\
#4 & #5 & #6 \\
#7 & #8 & #9
\end{array} \!\!\right)}

\newcommand{\sr}[1]{\stackrel{#1}{=}}
\newcommand{\mc}[1]{\multicolumn{1}{c}{#1}}
\newcommand{\MC}[3]{\multicolumn{#1}{#2}{#3}}
\newcommand{\TC}[4]{\contentsline {#1}{\numberline {#2}#3}{#4}}
\newcommand{\LIST}[3]{\contentsline {section}{\numberline {\hspace*{-0.5cm}#1}#2}{#3}}

\newcommand{\Hpi}[1]{\hat{\pi}_{_{\tiny\rm #1}}}
\newcommand{\titem}[1]{\vspace*{-0.15cm}\item[{\rm #1} ]\hspace*{0.1cm}}
\newcommand{\ub}[1]{\underline{\bf #1}}
\newcommand{\RED}[1]{{\color{red} #1}}

\newcommand{\two}[2]{{#1}_{_{\tiny\rm #2}}}
\newcommand{\three}[3]{{#1}_{_{#2, \tiny\rm #3}}}

\newcommand{\threev}[3]{\left(\!\!\begin{array}{c} #1 \\   #2 \\   #3 \end{array}
                          \!\!\right)}


\newcommand{\0}{{\bf 0}\!\!\!{\bf 0}}
\newcommand{\1}{{\bf 1}\!\!\!{\bf 1}}        \def\b1{{\bf 1\!\!\!1}}  


\newcommand{\ba}{{\bf a}}
\newcommand{\bb}{{\bf b}}
\newcommand{\bd}{{\bf d}}
\newcommand{\bn}{{\bf n}}
\newcommand{\bp}{{\bf p}}
\newcommand{\bq}{{\bf q}}
\newcommand{\br}{{\bf r}}
\newcommand{\bs}{{\bf s}}
\newcommand{\bt}{{\bf t}}
\newcommand{\bu}{{\bf u}}
\newcommand{\bv}{{\bf v}}
\newcommand{\bw}{{\bf w}}
\newcommand{\bx}{{\bf x}}
\newcommand{\by}{{\bf y}}
\newcommand{\bz}{{\bf z}}

\newcommand{\bA}{{\bf A}}
\newcommand{\bB}{{\bf B}}
\newcommand{\bC}{{\bf C}}
\newcommand{\bD}{{\bf D}}
\newcommand{\bH}{{\bf H}}
\newcommand{\bI}{{\bf I}}
\newcommand{\bL}{{\bf L}}
\newcommand{\bO}{{\bf O}}
\newcommand{\bP}{{\bf P}}
\newcommand{\bQ}{{\bf Q}}
\newcommand{\bS}{{\bf S}}
\newcommand{\bT}{{\bf T}}
\newcommand{\bU}{{\bf U}}
\newcommand{\bW}{{\bf W}}
\newcommand{\bX}{{\bf X}}
\newcommand{\bY}{{\bf Y}}
\newcommand{\bZ}{{\bf Z}}


\newcommand{\ibe}{{\bfm e}}
\newcommand{\ibp}{{\bfm p}}
\newcommand{\ibt}{{\bfm t}}
\newcommand{\ibv}{{\bfm v}}
\newcommand{\ibx}{{\bfm x}}
\newcommand{\iby}{{\bfm y}}
\newcommand{\ibz}{{\bfm z}}


\newcommand{\hb}{\hat{b}}
\newcommand{\hm}{\hat{m}}
\newcommand{\hp}{\hat{p}}
\newcommand{\hq}{\hat{q}}
\newcommand{\hR}{\hat{R}}

\newcommand{\hbr}{\hat{\br}}

\newcommand{\hth}{\hat{\theta}}   \newcommand{\hTh}{\hat{\Theta}}
\newcommand{\hbth}{\hat{{\bfm \theta}}}
\newcommand{\hvth}{\hat{\vartheta}}
\newcommand{\hmu}{\hat{\mu}}
\newcommand{\hsi}{\hat{\sigma}}   \newcommand{\hSi}{\hat{\Sigma}}
\newcommand{\hal}{\hat{\alpha}}   \newcommand{\hbe}{\hat{\beta}}
\newcommand{\hga}{\hat{\gamma}}
\newcommand{\hpsi}{\hat{\psi}}
\newcommand{\hxi}{\hat{\xi}}
\newcommand{\hpi}{\hat{\pi}}
\newcommand{\hla}{\hat{\lambda}}  \newcommand{\hbla}{\hat{{\bfm \lambda}}}

\newcommand{\whVar}{\widehat{\mbox{Var}}}
\newcommand{\whse}{\widehat{\mbox{se}}}
\newcommand{\whPr}{\widehat{\Pr}}

\newcommand{\tD}{\tilde{D}}
\newcommand{\tx}{\tilde{x}}
\newcommand{\ty}{\tilde{y}}

\newcommand{\tth}{\tilde{\theta}}     \newcommand{\tbth}{\tilde{{\bfm \theta}}}
\newcommand{\tpi}{\tilde{\pi}}
\newcommand{\tmu}{\tilde{\mu}}
\newcommand{\tbe}{\tilde{\beta}}
\newcommand{\tsi}{\tilde{\sigma}}     \newcommand{\tSi}{\tilde{\Sigma}}
\newcommand{\tpsi}{\tilde{\psi}}
\newcommand{\txi}{\tilde{\xi}}

\newcommand{\Bu}{\bar{u}}
\newcommand{\Bv}{\bar{v}}
\newcommand{\Bw}{\bar{w}}
\newcommand{\Bx}{\bar{x}}
\newcommand{\By}{\bar{y}}       \newcommand{\BY}{\bar{Y}}
\newcommand{\Bz}{\bar{z}}

\newcommand{\Bxi}{\bar{\xi}}

\newcommand{\BVar}{\overline{\mbox{Var}}}


\newcommand{\al}{\alpha}
\newcommand{\be}{\beta}
\newcommand{\ga}{\gamma}            
\newcommand{\Ga}{\Gamma}
\newcommand{\de}{\delta}            
\newcommand{\De}{\Delta}
\newcommand{\la}{\lambda}           
\newcommand{\La}{\Lambda}          
\newcommand{\Th}{\Theta}
\newcommand{\thx}{\theta_x}         \newcommand{\Thx}{\Theta_x}
\newcommand{\thy}{\theta_y}

\newcommand{\si}{\sigma}            \newcommand{\Si}{\Sigma}
\newcommand{\ka}{\kappa}
\newcommand{\om}{\omega}            \newcommand{\Om}{\Omega}

\newcommand{\ve}{\varepsilon}
\newcommand{\vp}{\varphi}
\newcommand{\vr}{\varrho}
\newcommand{\vth}{\vartheta}


\newcommand{\bxi}{{\bfm \xi}}      \newcommand{\bet}{{\bfm \eta}}
\newcommand{\bphi}{{\bfm \phi}}
\newcommand{\bmu}{{\bfm \mu}}      \newcommand{\bnu}{{\bfm \nu}}
\newcommand{\bla}{{\bfm \lambda}}  \newcommand{\bLa}{{\bfm \Lambda}}
\newcommand{\bSi}{{\bfm \Sigma}}
\newcommand{\bom}{{\bfm \omega}}   \newcommand{\bOm}{{\bfm \Omega}}
\newcommand{\bde}{{\bfm \delta}}   \newcommand{\bDe}{{\bfm \Delta}}
\newcommand{\bth}{{\bfm \theta}}
\newcommand{\bsi}{{\bfm \sigma}}
\newcommand{\bbe}{{\bfm \beta}}
\newcommand{\bpsi}{{\bfm \psi}}
\newcommand{\bpi}{{\bfm \pi}}


\newcommand{\Bernoulli}{\mbox{Bernoulli}}
\newcommand{\Binomial}{\mbox{Binomial}}
\newcommand{\BBinomial}{\mbox{BBinomial}}
\newcommand{\D}{\mbox{D}}
\newcommand{\Degenerate}{\mbox{Degenerate}}
\newcommand{\DExponential}{\mbox{DExponential}}
\newcommand{\Dirichlet}{\mbox{Dirichlet}}
\newcommand{\DMultinomial}{\mbox{DMultinomial}}
\newcommand{\Exponential}{\mbox{Exponential}}
\newcommand{\FDiscrete}{\mbox{FDiscrete}}
\newcommand{\GD}{\mbox{GD}}
\newcommand{\GDirichlet}{\mbox{GDirichlet}}
\newcommand{\GLiouville}{\mbox{GLiouville}}
\newcommand{\GPoisson}{\mbox{GPoisson}}
\newcommand{\Hgeometric}{\mbox{Hgeometric}}
\newcommand{\HPP}{\mbox{HPP}}
\newcommand{\IBeta}{\mbox{IBeta}}
\newcommand{\Ichi}{\mbox{I}\raisebox{0.5ex}{$\chi$}}
\newcommand{\IGamma}{\mbox{IGamma}}
\newcommand{\IGaussian}{\mbox{IGaussian}}
\newcommand{\IWishart}{\mbox{IWishart}}
\newcommand{\Laplace}{\mbox{Laplace}}
\newcommand{\Liouville}{\mbox{Liouville}}
\newcommand{\Logistic}{\mbox{Logistic}}
\newcommand{\Lognormal}{\mbox{Lognormal}}
\newcommand{\mBeta}{\mbox{Beta}}
\newcommand{\mGamma}{\mbox{Gamma}}
\newcommand{\Multinomial}{\mbox{Multinomial}}
\newcommand{\NBinomial}{\mbox{NBinomial}}
\newcommand{\ND}{\mbox{ND}}
\newcommand{\NDirichlet}{\mbox{NDirichlet}}
\newcommand{\NHPP}{\mbox{NHPP}}
\newcommand{\OIP}{\mbox{OIP}}
\newcommand{\OTP}{\mbox{OTP}}
\newcommand{\Poisson}{\mbox{Poisson}}
\newcommand{\TBeta}{\mbox{TBeta}}
\newcommand{\Wishart}{\mbox{Wishart}}
\newcommand{\ZIP}{\mbox{ZIP}}
\newcommand{\ZOIP}{\mbox{ZOIP}}
\newcommand{\ZTP}{\mbox{ZTP}}


\newcommand{\Corr}{\mbox{Corr}}
\newcommand{\Cov}{\mbox{Cov}}
\newcommand{\se}{\mbox{se}}
\newcommand{\Se}{\mbox{Se}}
\newcommand{\SE}{\mbox{SE}}
\newcommand{\tr}{\mbox{$\,$tr$\,$}}
\newcommand{\rank}{\mbox{rank}\,}
\newcommand{\Var}{\mbox{Var}}
\newcommand{\MSE}{\mbox{MSE}}
\newcommand{\CV}{\mbox{CV}}
\newcommand{\median}{\mbox{median}}

\newcommand{\logit}{\mbox{logit}}
\newcommand{\diag}{\mbox{diag}}
\newcommand{\data}{\mbox{data}}
\newcommand{\KL}{\mbox{KL}}
\newcommand{\IG}{\mbox{IG}}
\newcommand{\I}{\mbox{I}}

\newcommand{\qand}{\quad \mbox{and} \quad}
\newcommand{\qas}{\quad \mbox{as} \quad}
\newcommand{\qag}{\quad \mbox{against} \quad}
\newcommand{\qor}{\quad \mbox{or} \quad}
\newcommand{\qve}{\quad \mbox{versus} \quad}
\newcommand{\col}{\mbox{: }}
\newcommand{\RE}{\mbox{RE}}
\newcommand{\yes}{\mbox{yes}}
\newcommand{\No}{\mbox{no}}
\newcommand{\DPP}{\mbox{DPP}}

\newcommand{\e}{\mbox{e}}
\newcommand{\w}{\mbox{w}}
\renewcommand{\ge}{\geqslant}
\renewcommand{\le}{\leqslant}


\newcommand{\II}{I$\!$I}
\newcommand{\III}{I$\!$I$\!$I}
\newcommand{\IR}{{I\!\! R}}
\newcommand{\IV}{I$\!$V}
\newcommand{\et}{{\it et al}.}
\newcommand{\Et}{{\it et al}.\,}

\newcommand{\yikH}{y_{ik}^{\rm H}}
\newcommand{\ybH}{\bar{y}^{\rm H}}


\newcommand{\sd}{\stackrel{{\rm d}}{=}}
\newcommand{\sdt}{$ {\small $\sd$} $}
\newcommand{\heq}{\,\hat{=}\,}
\newcommand{\iid}{\stackrel{{\rm iid}}{\sim}}
\newcommand{\tiid}{i.i.d.$\hspace*{0.08cm}$}
\newcommand{\ind}{\stackrel{{\rm ind}}{\sim}}
\newcommand{\dsim}{\stackrel{.}{\sim}}
\newcommand{\dis}{\displaystyle}
\newcommand{\tex}{\textstyle}
\newcommand{\cf}{cf.$\hspace*{0.1cm}$}

\newcommand{\T}{\!\top\!}
\newcommand{\na}{\nabla}
\newcommand{\noi}{\noindent}
\newcommand{\ra}{\rightarrow}
\newcommand{\pr}{\propto}
\newcommand{\eq}{\equiv}
\newcommand{\pa}{\partial}
\newcommand{\ol}{\overline}
\newcommand{\non}{\nonumber}
\newcommand{\ap}{\approx}
\newcommand{\Bot}{\;\bot\;}
\newcommand{\inde}{{\Bot\!\!\!\!\!\!\!\Bot}}
\newcommand{\btu}{\bigtriangleup}


\newcommand{\vs}{\vspace*{-0.25cm}}
\newcommand{\vkl}{\vskip 0.10in}
\newcommand{\vkL}{\vskip 0.15in}
\newcommand{\vkU}{\vskip 0.30in}


\newcommand{\namelistlabel}[1]{\mbox{#1}\hfil}
\newenvironment{namelist}[1]{%
\begin{list}{}
       {\let \makelabel \namelistlabel
        \settowidth{\labelwidth}{#1}
        \setlength{\leftmargin}{1.1\labelwidth}   }
        }{%
\end{list} }

\def\bds{\begin{description} \itemsep=-\parsep \itemindent=-0.7 cm}
\def\eds{\end{description}}
\def\i{\item}

\newcommand{\hphi}{\hat{\phi}}
\newcommand{\tphi}{\tilde{\phi}}
\newcommand{\tla}{\tilde{\lambda}}
\newcommand{\hhom}{\hat{\om}}
\newcommand{\htau}{\hat{\tau}}
\newcommand{\hrho}{\hat{\rho}}



\baselineskip 0.25in \vskip 0.05in \noi {\bf Abstract}. Visualization of extremely large datasets in static or dynamic form is a huge challenge because most traditional methods cannot deal with big-data problems. A new visualization method for big-data is proposed based on Projection Pursuit, Guided Tour and Data Nuggets methods, that will help display interesting hidden structures such as clusters, outliers and other nonlinear structures in big-data. The Guided Tour is a dynamic graphical tool for high-dimensional data combining Projection Pursuit and Grand Tour methods. It displays a dynamic sequence of low-dimensional projections obtained by using Projection Pursuit (PP) index functions to navigate the data space. Different PP indices have been developed to detect interesting structures of multivariate data but there are computational problems for big-data using the original guided tour with these indices. A new PP index is developed to be computable for big-data, with the help of a data compression method called “Data Nuggets” that reduces large datasets while maintaining the original data structure. Simulation studies are conducted and a real large dataset is used to illustrate the proposed methodology. Static and dynamic graphical tools for big-data can be developed based on the proposed PP index to detect nonlinear structures.

\vskip 0.10in \noi {\bf Keywords}: Big Data; Projection Pursuit; Guided Tour; Data Nuggets.

\baselineskip 0.30in
\setcounter{equation}{0}
\section{$\!\!\!\!\!\!\!$. Introduction}  
In today’s world, huge amounts of data are continuously stored in almost all fields. This creates many computational and methodological issues when analyzing big data with standard methods and tools. For example, data visualization plays an important role in displaying the essential information and structures of data, that maybe useful in business decision making, scientific finding, etc. But it's a huge challenge to visualize extremely large data sets in static or dynamic form because most traditional data visualization methods can't deal with big-data problems. The aim of our research is to develop new visualization methods for big-data, based on a combination of Projection Pursuit, Guided Tour and Data Nuggets methods, that will help display interesting hidden structures such as clusters, outliers and other nonlinear structures.

A common visualization technique to explore multivariate data is Projection Pursuit. It is an effective method for finding static low-dimensional projections that uncover interesting structures, via the optimization of Projection Pursuit (PP) indices. A PP index function numerically measures features of low-dimensional projections for which higher values correspond to more "interesting" structures, such as point mass, holes, clusters, outliers and other linear and nonlinear structures. There are two kinds of PP indices. One is for finding specific structures such as Holes Index, Central Mass Index and Linear Discriminant Analysis Index, etc.. For instance, the Holes index could help find hole structures in the projected data, such as two separate clusters for one-dimensional projections. The central mass structures and outliers could be found by CM index, and LDA index could find multiple clusters on the projected data. The other kind of PP indices we are focusing on is for finding general non-linear structures, including the Legendre Index (Friedman, 1987), the Hermite Index (Hall, 1989) and the Natural Hermite Index (Cook, 1993). These three indices were all developed to detect departures from normality because unusual projections look non-normal. As the dimension goes to infinity, by CLT, any random one-dimensional projection will be a linear combination of all the variables, and therefore it would be approximately normally distributed. This result also holds for two dimensional and other low-dimensional projections. Therefore, if there is a significant difference between projected data and normal distribution, some non-linear structures would be shown from the projected data. Also, PP indices could be optimized to find the static most unusual and "interesting" low dimensional projection, which indicates the hidden structure of high-dimensional data.

Grand Tour is a technique of data visualization providing a dynamic sequence of interpolated random low dimensional projections, to visually extrapolate from low-dimensional shapes to multidimensional structures. Projection Pursuit and Grand Tour are combined into a dynamic graphical tool for exploring high-dimensional data, called a Projection Pursuit Guided Tour (Cook, 1995). It uses a PP index function to navigate the data space, so the next projection plane during the tour is selected by optimizing a PP index. The PP guided tour is available in the R package, \emph{tourr} (Wickham et al., 2011), which provides optimization routines and a data visualization tool. 

However, the original guided tour is computationally expensive for visualizing big-data and most data visualization tools including \emph{tourr} ran into computational problems with big-data. In this paper, for the purpose of visualizing big-data in a more efficient way, we propose a new Projection Pursuit index that can be computable for big-data, with the help of a data compression method called “data-nuggets” that reduces a large dataset into a smaller collection of data nuggets that maintain the overall data structure. The new PP index can be used to develop new data visualization tools including a guided tour to generate interactive, dynamic and efficient visualization for big-data. In section 2, the methodology of the new PP index for big-data is proposed. Section 3 shows the performance of the new index on simulated data and optimization of the index is implemented in section 4. A real data set is used to illustrate the proposed methodology in section 5 and future work is discussed in section 6. 

\section{$\!\!\!\!\!\!\!$. A new Projection Pursuit index for big data}  

Assume that $\mathbf{X}\in \mathbb{R} ^{{{n}}\times {{p}}}$ be a big data matrix representing a $p$-dimensional dataset, and assume that X has been spherized by a linear transformation. Let $A \in \mathbb{R}^{{{p}}\times {{d}}}$ be a random orthogonal projection matrix that projects $\mathbf{X}$ into a $d$-dimensional space ($d<p$). Let $\mathbf{Y}=\left[\mathbf{y}_{1}, \mathbf{y}_{2}, \cdots, \mathbf{y}_{n}\right]^{T}\heq\mathbf{X A}  \in \mathbb{R} ^{{{n}}\times {{d}}}$ be the projected data in the $d$-dimensional space. Consider density $f(\mathbf{y})$ of the projected data $\mathbf{Y}$ and standard multivariate Gaussian density $\phi(\mathbf{y})$. An index, $I,$ can be constructed by measuring the distance of $f(\mathbf{y})$ from the standard normal. The Legendre index, $I^{L}$, proposed by Friedman (1987) is defined by
$$
{I}^{L}=\int_{\mathbb{R}^d}\{f(\mathbf{y})-\phi(\mathbf{y})\}^{2} \frac{1}{2 \phi(\mathbf{y})}d\mathbf{y}
$$
Observing upweighted tails in the Legendre index, Hall (1989) proposed the Hermite index that measures the $L^{2}$ -distance between $f(\mathbf{y})$ and the standard normal density with respect to Lebesgue measure:
$$
I^{H}=\int_{\mathbb{R}^d}\{f(\mathbf{y})-\phi(\mathbf{y})\}^{2} d\mathbf{y}
$$
Based on the Hermite Index, Cook (1993) proposed the Natural Hermite Index:
$$
  I^{N}=\int_{\mathbb{R}^d}\{f(\mathbf{y})-\phi(\mathbf{y})\}^{2} \phi(\mathbf{y}) d \mathbf{y}  
$$
It can be regarded as a weighted version of the Hermite Index and it's simple and more radical in its treatment of tail weight. Therefore, here we consider using the same idea of the Nature Hermite Index to deal with big data. 

When dealing with extremely large data sets, it's computationally expensive for those PP indices because of the density estimation and calculation of the integral. Moreover, guided tour is hard to implement for big data because the index need to be optimized at each step during the tour, which will take much longer time with large data sets. Therefore, we propose a new PP index based on the data nugget method that was developed by Beavers (2019) for reducing large datasets while maintaining the data structure.

The data nuggets method reduces a large dataset into a smaller collection of nuggets of data while maintaining the overall structure of the original data. Each nugget corresponds to a subset of the data represented by its center, weight, and scale parameters. The center and weight of a nugget are the mean and the number of observations inside the nugget respectively. The scale $s_{i}= \sqrt{\operatorname{Max}\{\operatorname{diag}\left(\operatorname{Cov}\left(\mathbf{X}_{i}\right)\right)\}}$ where $\mathbf{X}_{i}$ is the observations belonging to data nugget $i$. When the weight of data nugget is 1, the scale $s_{i}=0$. Generally, the number of data nuggets would depend on the complexity of structure of data, but for computational purposes, it's preferred to construct 1000-10000 data nuggets from the data.  Data nuggets that are not spherical can be refined by splitting them when the eigenvalues of the nugget's covariance matrix are not similar. After refining a set of data nuggets, they would be more spherical than the original set. Ideally the data nuggets should be spherical with similar radius.

In order to implement projection pursuit for big data, we convert big data into a set of a few thousands data nuggets. The new PP index for big data is constructed based on the obtained refined data nuggets using the idea of the Natural Hermite Index. The construction process and its definition are described below.\\
\textbf{Definition 1} The new Natural Hermite Index for big data is constructed with data nuggets as follows:\\
\emph{Step 1.} Construct and refine data nuggets from the original big data set to obtain $m$ data nuggets with centers, weights and scales.\\
\emph{Step 2.} Spherize the set of data nugget centers with weights, i.e., weighted mean is zero and weighted covariance is identity matrix.\\
\emph{Step 3.} Let $\mathbf{X}\in \mathbb{R} ^{{{m}}\times {{p}}}$ denote the nugget centers spherized with weights $\mathbf{w}$ where $\mathbf{w}\in \mathbb{R}^{m} $ is the weight vector of the data nuggets. Let $A \in \mathbb{R}^{{{p}}\times {{d}}}$ be an orthogonal projection matrix. Let $\mathbf{Y}\heq \mathbf{X A}=\left[\mathbf{y}_{1}, \mathbf{y}_{2}, \cdots, \mathbf{y}_{m}\right]^{T}\in \mathbb{R} ^{{{m}}\times {{d}}}$ be the projected nugget centers and  let $\mathbf{s}\in \mathbb{R}^{m} $ be the scale vector. A new Natural Hermite Index based on the data nuggets is defined by
\begin{equation}
I^B = \int_{\mathbb{R}^d}\{\hat{f}_B(\mathbf{y})-\phi(\mathbf{y})\}^{2} \phi(\mathbf{y}) d \mathbf{y}    
\end{equation}
where $\mathbf{y}\in \mathbb{R}^d $, $\phi(\cdot)$ is the standard multivariate normal density, and the estimated density function is calculated based on the nuggets:
\begin{equation}
\hat{f}_B(\mathbf{y}) = \sum_{i=1}^{m} \frac{w_i}{\sum_{i=1}^{m}w_i}|\mathbf{S_i}|^{-1/2} \phi\left(\mathbf{S_i}^{-1/2} (\mathbf{y}-\mathbf{y}_{i})\right)   
\end{equation}
where $\mathbf{S_i} = \operatorname{max}\{s_i^2,\delta\} \mathbf{I}_d$ with a pre-determined minimal scale level $\delta$. Alternatively, $\mathbf{S_i} = \operatorname{median}(\mathbf{s}^2) \mathbf{I}_d$ for all $i = 1, 2, ..., m$ if $\mathbf{s}$ has a wide range.

Here, for large data sets, the density of projected data $\hat{f}_B(\mathbf{y})$ is estimated based on obtained refined data nuggets. It's calculated by the multivariate kernel density estimation but considering a weighted sum rather than mean values, with weights and scales of data nuggets. Ideally, the raw large data set should be spherized first before constructing data nuggets because projection pursuit is performed on the spherized data. But it would have computational limitations if the size of data set is large. To reduce the computational efforts, the proposed process in Definition 1 only conducts the spherization on the constructed and refined data nuggets with their weights. It helps obtain approximate solutions in a concise way, especially for large data sets. Additionally, it is recommended that the big data set is standardized first before constructing data nuggets, especially if the scales of variables are diverse.

This new Natural Hermite Index is computationally much cheaper than using the original PP indices for the whole data set, because calculations are based on the obtained smaller collection of data nuggets rather than the large data set. Also, since data nuggets maintain the overall data structure, this new Natural Hermite Index will help find hidden structures in the original big data set.

\section{$\!\!\!\!\!\!\!$. Performance of the new index on simulated data}  

In order to check the performance of the proposed PP index for big data, we conducted simulations with simple data sets that we generated. To generate the data we first constructed the data nugget centers equally spaced on the grid and provide them with random weights and scales. After obtaining all the nuggets, we generate raw data from a uniform distribution inside a sphere of radius $2s$ around each data nugget center, where $s$ is the scale parameter. Then the original Natural Hermite Index $I^N$ for the raw data and the new PP index $I^B$ based on the data nuggets are calculated and compared. We conducted simulations for dimensions $p = 4$ and $p = 6$, and considered projections with dimension $d = 2$. Moreover, in order to make the raw data with a hidden structure, the first variables of nugget centers were adjusted to make two obvious clusters in the simulated data. 

Firstly, for $p = 4$, we constructed 256 data nuggets with centers equally spaced on $[-2,2]^4$ and 9106 observations of raw data are generated. Figure 1 shows the histograms of the first variables of the raw data and nugget centers, from which two obvious clusters are shown. Figure 2 shows two examples of $2$-dimensional random projections. The left ones are the projected raw data and right ones are the projected data nuggets with the same projection matrix. The color of each data nugget corresponds to its weight, which is the number of observations in the nugget. Lighter green means a larger weight while darker green means a lower weight. For Figure 2(a), there's no obvious structure and the original Natural Hermite Index $I^N$ for the raw data is \textbf{0.000128} while the new PP index $I^B$ equals \textbf{0.000216}. For Figure 2(b),  there are two clear clusters and $I^N$ for the raw data is \textbf{0.008222} while the new PP index $I^B$ based on the data nuggets equals \textbf{0.007861}. Therefore, the values of the new PP index based on data nuggets are very similar with the original index for the whole raw data. 
\begin{figure}[H]
		\centering
		\includegraphics[width=4.7in]{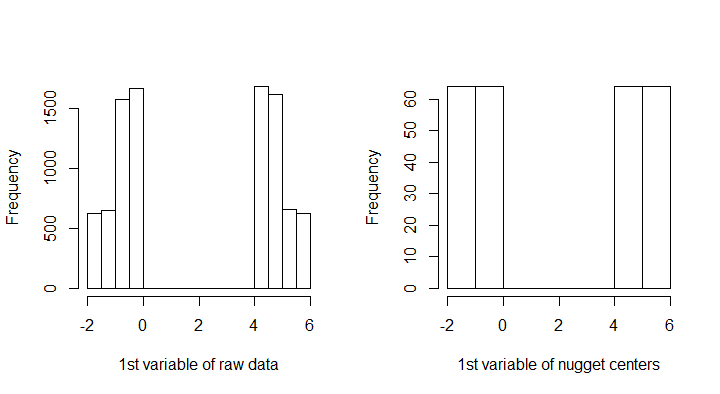}
		\caption{\small \it Histograms of the first variables of the raw data and data nuggets for 4-dimensional simulated data}
\end{figure}
\begin{figure}[H]
		\centering
		\includegraphics[width=4.3in,height = 3.2in]{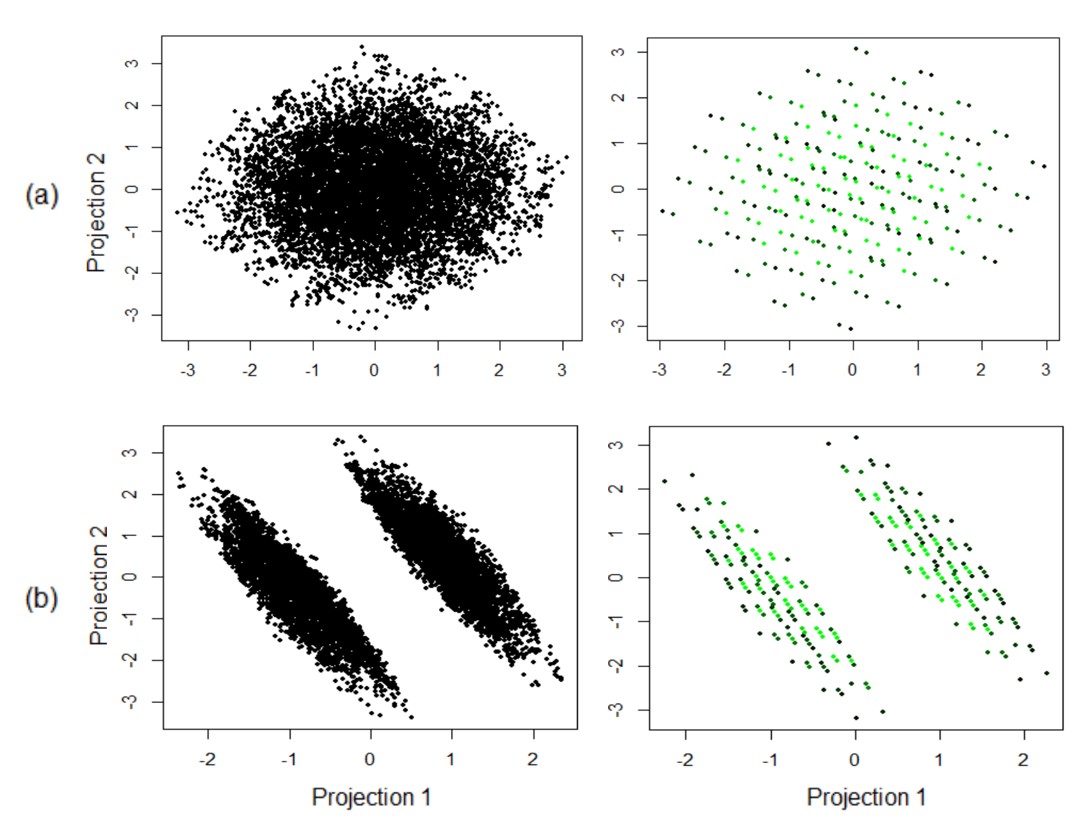}
		\caption{\small \it 2-dimensional random projected raw data and data nuggets of the 4-dimensional simulated data}
\end{figure}

In order to check the performance of the new PP index for finding interesting projections, we formed a sequence of projections by iterating a small random rotation $R_\delta$ over the projection in Figure 2(b). We completed the path projections in the opposite direction by iterating $R_\delta^{-1}$. In this way we constructed a path of projections that follows a geodesic on the manifold of projections and is centered around the projection in Figure 2(b). Figure 3 shows the path of rotations plus the two curves of the indices, the top red one for the new PP index $I^B$ based on data nuggets and the bottom blue one for the original Natural Hermite Index $I^N$ for the raw data. The index values correspond to the projections at the middle. The plots shown between the two curves are the projected data nuggets and raw data for each rotated projections. The highlighted middle plot of the sequence shows the  same projection as Figure 2(b) clearly indicating two clusters. We can find that the projected structure is getting blurry on both directions. Therefore, in this local region, by rotating the projection plane a little bit, we can find the local optimal projections found by the new PP index and the original Natural Hermite Index are the same. Also, the values and trends of the two indices are similar to each other. 
\begin{figure}[h]
		\centering
		\includegraphics[width=6.5in,height = 4.5in]{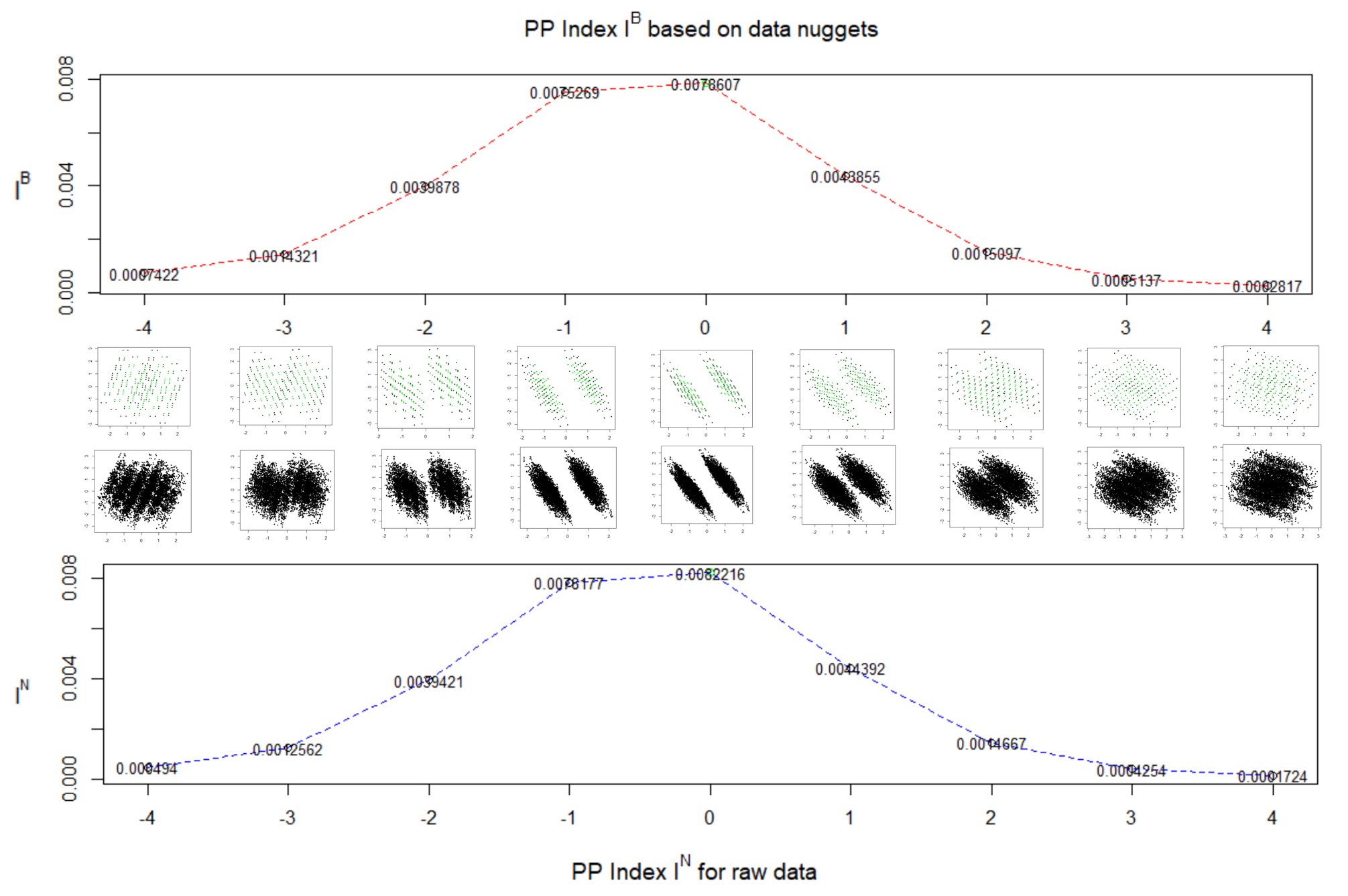}
		\caption{\small \it The path of rotations and two curves of $I^B$, $I^N$ for the 4-dimensional simulated data. The x axis represents different projections.}
\end{figure}

We also conducted similar simulations for $p = 6$. There are 88218 raw data with 4096 data nuggets. Figure 4 displays two $2$-dimensional projections showing a random view of the data and a view of the underlying structure, respectively. Figure 4(a) shows no obvious structure and $I^N$ for the raw data is \textbf{0.000098} while the new PP index $I^B$ equals \textbf{0.000107}. For Figure 4(b) showing two clear clusters, the new PP index $I^B$ based on the nuggets equals \textbf{0.003067} while $I^N$ for the raw data is \textbf{0.003661}.

Similarly, based on the projection in Figure 4(b), we iterated a small rotation on both directions and obtained the path of projections in Figure 5. Figure 5 displays two curves of the Natural Hermite indices for the projections of the original data and the data nuggets respectively. The graph shows a high similarity of the two curves in both the trends and the local optimum.
\begin{figure}[h]
		\centering
		\includegraphics[width=4.3in,height = 3.2in]{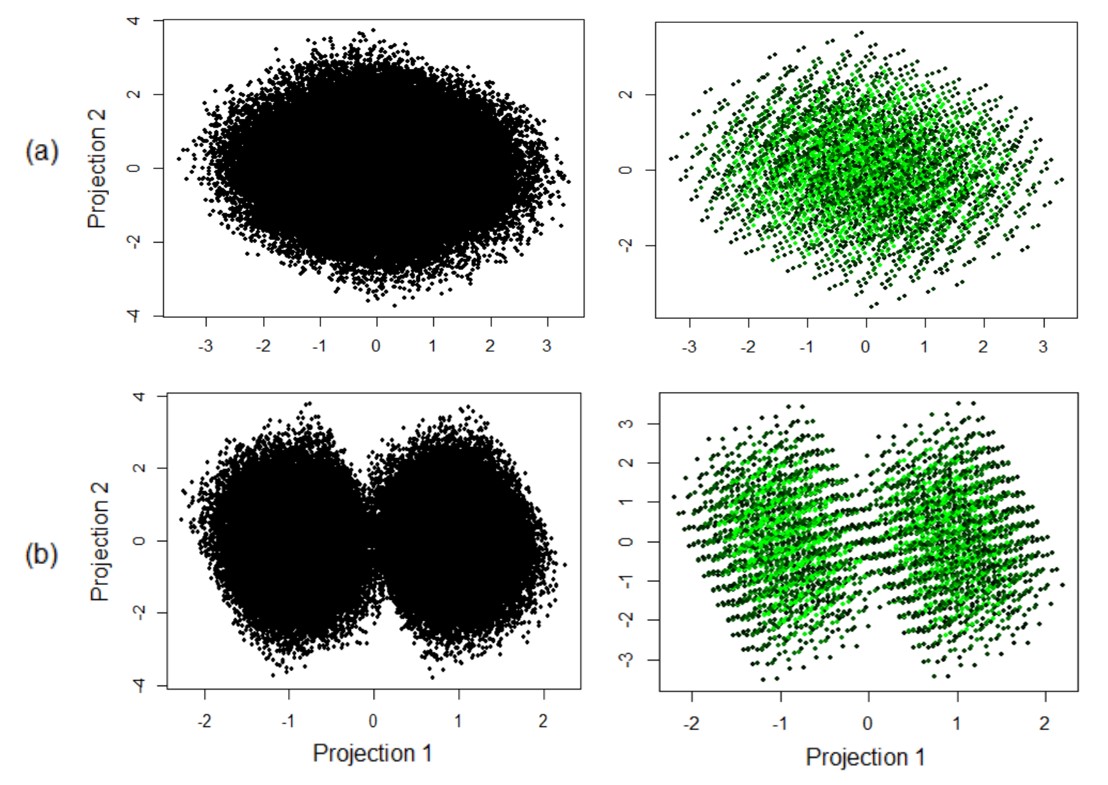}
		\caption{\small \it 2-dimensional random projected raw data and data nuggets of the 6-dimensional simulated data}
	\end{figure}
\begin{figure}[h]
		\centering
		\includegraphics[width=5.3in,height = 2.7in]{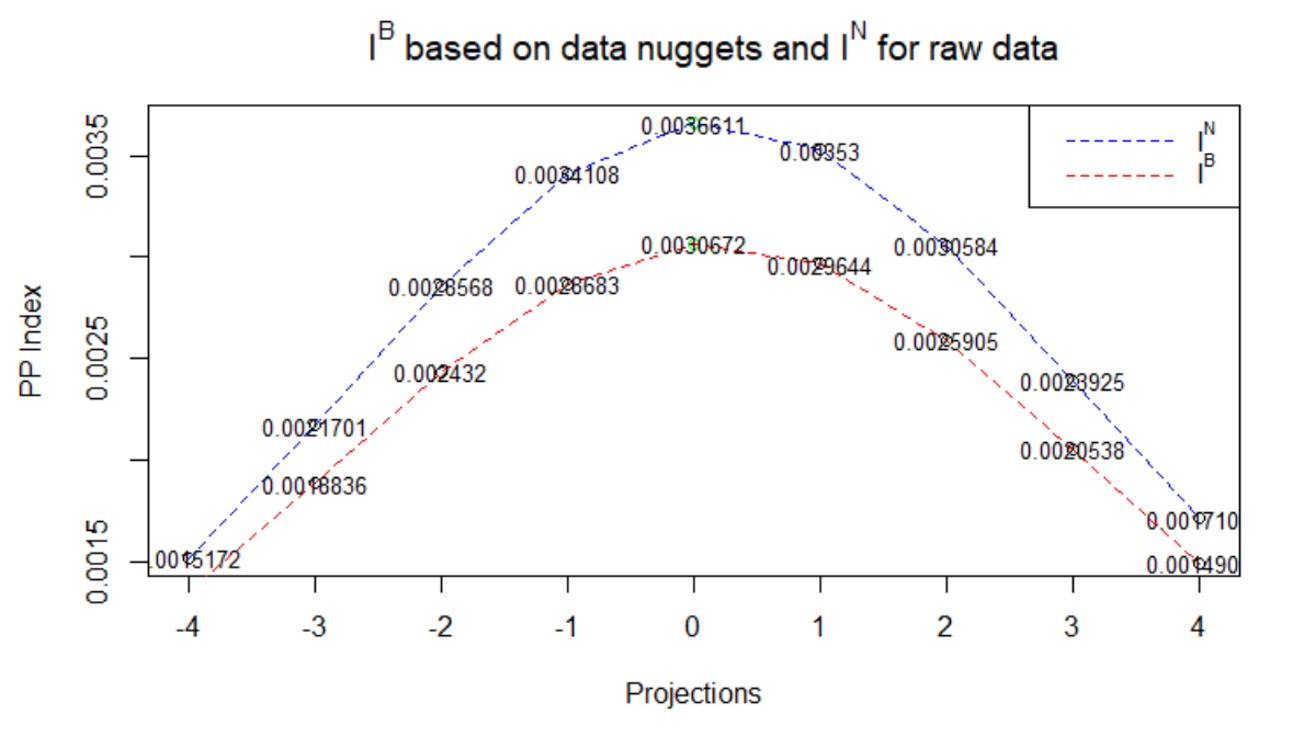}
		\caption{\small \it PP indices along the path of rotations for the 6-dimensional simulated data}
\end{figure}

The calculation of the original Natural Hermite index for the projected whole data involves $n$ evaluations of each of $n$ densities centered at each point so it is of order $n^2$. For the projected data nuggets the evaluation of the new PP index is of the order $m^2$. The projection calculation is of linear order so the computation of the two indices is of order $n^2$ and $m^2$ respectively. In the examples of Figure 2, since $m=256$ and $n=9106$, the savings by using data nuggets is of the order $n^2/m^2 \sim 10^3$.  For the example in Figure 4 the savings is a factor of $10^2$.
The results above show that the performance of the new PP index with the data nuggets on simulated data is comparable to the original index on the full data but the computation is of an order of magnitude less.

\section{$\!\!\!\!\!\!\!$. Optimization of the new index}  

Apart from the performance of the new PP index itself, we implemented the optimization algorithm for it to find the most interesting static projection which has the largest PP index value. Here we used the modified simulated annealing method that Lee et al. (2005) used, with a neighborhood temperature parameter and the other (cooling parameter) for the probability that guards against getting trapped in a local maximum. The neighborhood temperature is rescaled by the other cooling parameter enabling escape from the local maximum to look for other maxima. For the purpose of illustration, we used the optimization algorithm with the cooling parameter and initial temperature both equaling 0.9, for the new PP index $I^B$ and the Natural Hermite index $I^N$. We applied it on the two simulated data sets we analyzed in Section 3 and compared the projections found by optimizing the indices. 

For the data nuggets, we implemented the algorithm for the new PP index $I^B$, and for the raw data we optimized the original Natural Hermite index $I^N$. For the 4-dimensional simulated data set of which two random projections are shown in Figure 2, the optimal new PP index we found for the nuggets equals \textbf{0.016647} while the optimal $I^N$ for the whole raw data is \textbf{0.013007}. The left and middle plots of Figure 6(a) are the optimal projection found by the new PP index $I^B$ for the nuggets. The left plot shows projected data nuggets and many nuggets are overlapped with each other. The middle plot of Figure 6(a) shows the projected raw data using the optimal projection matrix found by the new PP index $I^B$ for the nuggets, for which the estimated Natural Hermite index $I^N$ equals \textbf{0.012709}. The right plot of Figure 6(a) displays the optimal projection of the whole raw data found by the Natural Hermite index $I^N$ with a $2$-dimensional small-angle rotation. 
\begin{figure}[h]
		\centering
		\includegraphics[width=6.3in]{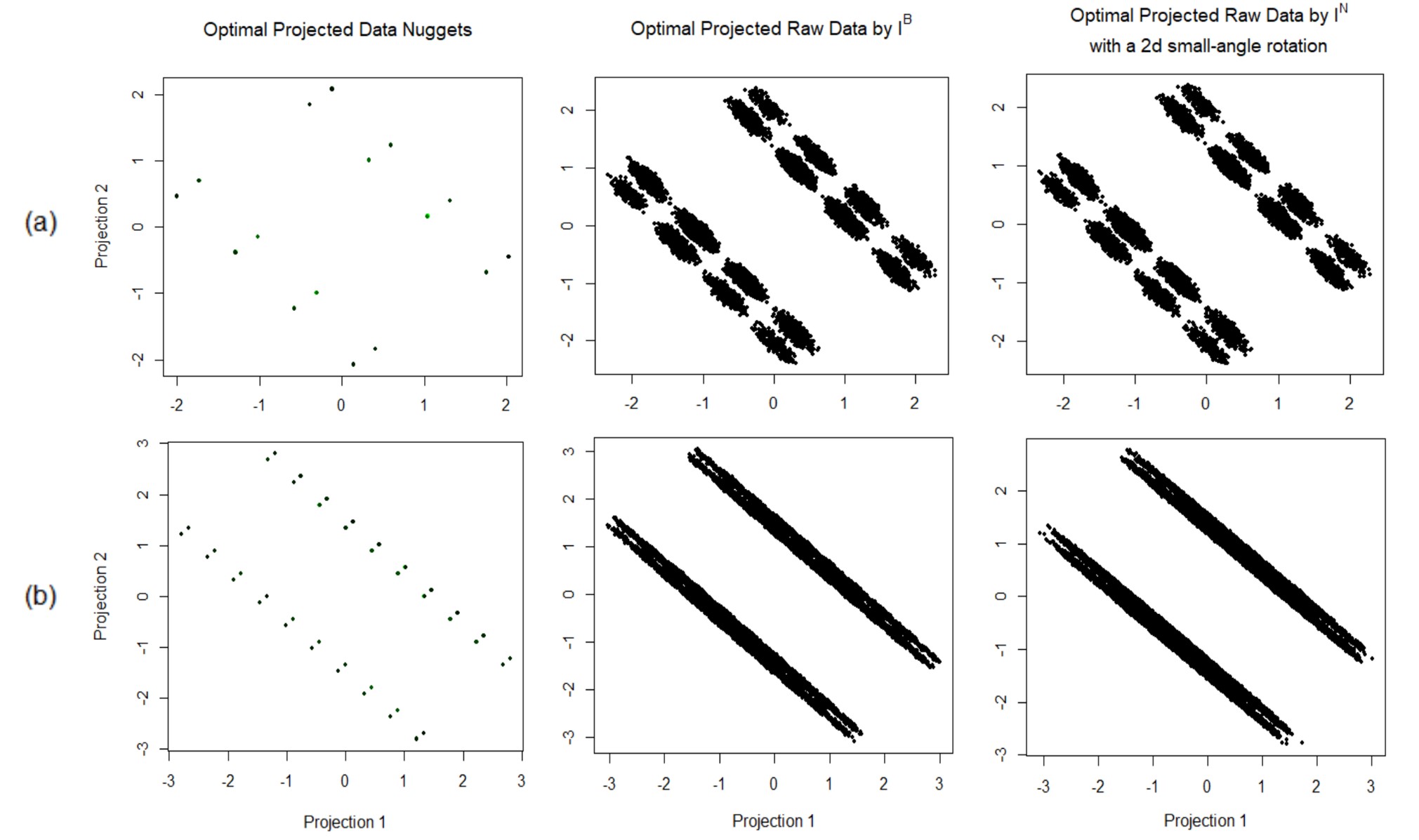}
		\caption{\small \it The optimal projections found for the (a) 4-dim simulated data and (b) 6-dim simulated data.}
\end{figure}

Similar results are shown in Figure 6(b) for the $6$-dimensional simulated data set for which Figure 4 shows two random projections. The optimal projection of the data nuggets found by maximizing $I^B$ is shown in the left plot of Figure 6(b) and the right plot is the slightly rotated optimal projection of the whole raw data by optimizing $I^N$. The optimal new PP index $I^B$ for the nuggets equals \textbf{0.012011} and the value of the optimal $I^N$ for the entire data set equals \textbf{0.019609}. For the projection shown in the middle plot of Figure 6(b), which is the projected raw data with the optimal projection matrix found by $I^B$, its estimated Natural Hermite index $I^N$ equals \textbf{0.019447}. 

For both simulated data sets, the optimal projections found by the new PP index for the data nuggets show clearly the underlying two clusters and they are approximately same with those found by the original Natural Hermite index for the entire data set but with much less computational cost.

\section{$\!\!\!\!\!\!\!$. A real data application to flow cytometry}  

Cell flow cytometry experiments are common in immunological research and they use fluorochrome-conjugated antibodies to differentiate cell populations based on surface and internal expression of delineating proteins. Data sets generated by a standard cell flow cytometry experiments routinely include the measurement of the abundance of 12 or more proteins on the surface of each cell. Among those proteins, three of them could be used to characterize B-cells. Beavers, Cabrera, et al. (2019) analyzed a flow cytometry dataset of an experiment involving 9 proteins measurements of 1 million B-cells. There are 1,048,575 B-cells with 9 intensity values of fluorochrome indicating abundance of 9 kinds of surface proteins. The objective of the experiment was to explore B-cell surface protein profiles that characterize B-cell states, specifically, to detect clusters of B-cells with a protein profile suggesting high activity modes and to estimate the proportion of highly-active B cells in the sample. Projection Pursuit is a good way to find hidden structures in it but calculating the original PP indices based on this whole large data set is computationally expensive and it's also hard to perform the guided tour for the raw data. Here we apply the proposed methodology to this big data set to explore interesting structures inside it. 

After obtaining 3431 refined data nuggets from this large high-dimensional dataset, we used weighted PCA to get the principal components of them. Figure 7(a) shows the projection of the 1st and 2nd principal components and the projection in Figure 7(b) is obtained by the 2nd and 3rd principal components. The left plots in Figure 7 are the projected data nuggets while the right plots are the corresponding projected raw data with the same projection matrices. The middle plots show the estimated densities with contours which are calculated by (2.2) with the weights and scales of data nuggets. And by (2.1), we obtain the new PP index values for these two principal components projections, which equal 0.030307 and 0.070785 for Figure 7(a) and (b) respectively. There's no obvious structure in Figure 7(a) with its small index value and Figure 7 (b) shows a vague L-shape with a little higher index. 

By checking $2$-dimensional random projections, we found two interesting projections shown in Figure 8(a) and (b). Figure 8(a) shows an approximately diamond shape with four clusters at the vertices and the most populated one is at the left vertex. For Figure 8(b), there is a vague clustering structure with a large cluster at the left side and the other small one on the right. Figure 8(c) and (d) are the projections with larger index values, and indeed (d) is the optimal $2$-dimensional projection we found by the optimization algorithm implemented in section 4 for the new PP Index $I^B$ based on data nuggets. The estimated new PP index values, $I^B$, for those four projections in Figure 8 are: (a) 0.033398, (b) 0.043588, (c) 0.144126 and (d) 0.164447. 

For Figure 8(c), it displays a Y-shape and looks like a spaceship, and the shape of the optimal projection in Figure 8(d) is like two letters F back-to-back together. Compared with the projections in Figure 7, Figure 8(a) and (b), the projections with much higher new PP indices in Figure 8(c) and (d) display significantly more clear unusual shapes, indicating potential data structures inside the large data set. In addition, we generated several random samples from the original large data set and used the original Natural Hermite index to find the optimal projections. There are 3000 data points in each sample. The optimal projections found using random samples are mostly like the L-shape shown in Figure 7(b) and a few of them show different shapes. For those projections, the minimal and maximal index values are 0.011664 and 0.020622 respectively, which are much smaller than the index value of the optimal projection found by our new method. Therefore, the proposed new PP index based on the data nuggets will help find more interesting projections and hidden structures in the whole large data set with considerably less computational cost than the original PP indices. And by Figure 8(a) and (b), it seems that a new PP index specifically for the clustering structure based on data nuggets can be developed with great prospects. 

For the purpose of finding the biological interpretations of the optimal projections, varimax rotations were conducted on the two optimal projections in Figure 8(c) and (d) respectively. For each projection, the orthogonal basis was rotated to express each of the two main axes of the projections as a function of a few original variables, i.e., surface proteins. The initial process of spherizing data nugget centers with weights is also considered with the optimal projection matrices during the varimax rotation in order to align with the original variables. Figure 9(a) and (b) are the rotated projections after varimax rotations from the two optimal projections in Figure 8(c) and (d). The variable loadings for each kind of surface protein are represented in the loadings plots shown in Figure 9(c) and (d), for the rotated projections in Figure 9(a) and (b) respectively. The true names of each surface protein are not included in the raw dataset because of data privacy issues, and instead, letters A-I are used to represent each protein. In the loadings plots, the proteins with large loadings that are bigger than 0.8 for the first component are colored blue, and the proteins colored red are those with large loadings for the second component. The proteins with both colors have significant loadings on both components. For both optimal projections in Figure 9(a) and (b), proteins C and G contribute strongly to the first component, and protein A to the second component. For the projection in Figure 9(a), protein G and protein I also have a strong relationship with the second component. Protein B contributes significantly to the second component for the projection in Figure 9(b). If only keeping the loadings of those significant proteins and making others zero, the overall structures in Figure 9(a) and (b) will be maintained for both projections. The findings may help scientists to reveal potential relationships between these surface proteins and B-cell states. 

In addition, in order to detect clusters of B-cells that can be characterized as highly active cells, which scientists show much interest in, weighted K-means was applied to the optimal 2-d projections. The rotated projections in Figure 9(a) and (b) were clustered by weighted K-means considering the weights of data nuggets. Ten initial centers were used and the cluster configuration with the least total weighted within-cluster sum of squares (WWCSS) was chosen. The number of clusters was chosen from 2 to 12 for each projection. The total weighted within-cluster sum of squares (WWCSS) was calculated for each number of clusters. The best number of clusters was chosen based on the ratio of the second difference of the total WWCSS to its first difference. For each number of cluster $k$, the second difference of total WWCSS, i.e., the difference between the decrease of total WWCSS from $k-1$ to $k$, and its decrease from $k$ to $k+1$, was calculated first. Then a ratio of the second difference to the first difference which is the decrease of total WWCSS from $k-1$ to $k$, was calculated and compared for different $k$s. The optimal number of clusters was chosen with the largest ratio between the second difference and the first difference of total WWCSS. For the rotated projection in Figure 9(a), eight clusters were chosen and for the one in Figure 9(b), the best number of clusters was nine. 

Figure 10(a) and Figure 11(a) show the clustering results for the two projections in Figure 9(a) and (b) respectively. The box plots given in Figure 10(b) and Figure 11(b) summarize the level of expression within each cluster for each protein. The box plots are constructed based on the values of data nugget centers before the process of spherizing. There are a few kinds of proteins showing significant differences in expression levels between clusters, and the patterns are variable for different kinds of proteins. For example, in Figure 11(b), Protein A, C, and G have distinct distributions of expression levels for different clusters. For those proteins that scientists are interested in, the differences in expression levels in clusters may reveal potential protein functions and profiles indicating B-cell states.

For the data about millions of B-cells with multiple protein measurements from cell flow cytometry experiments, the proposed PP index method based on data nuggets could help find optimal low-dimensional projections of data in a short time. The optimal projections could reveal potential hidden structures about protein profiles of B-cells,  which could help scientists investigate the surface protein functions, and obtain more information about relationships between different kinds of protein and B-cell states.

\begin{figure}[H]
		\centering
		\includegraphics[width=6.3in]{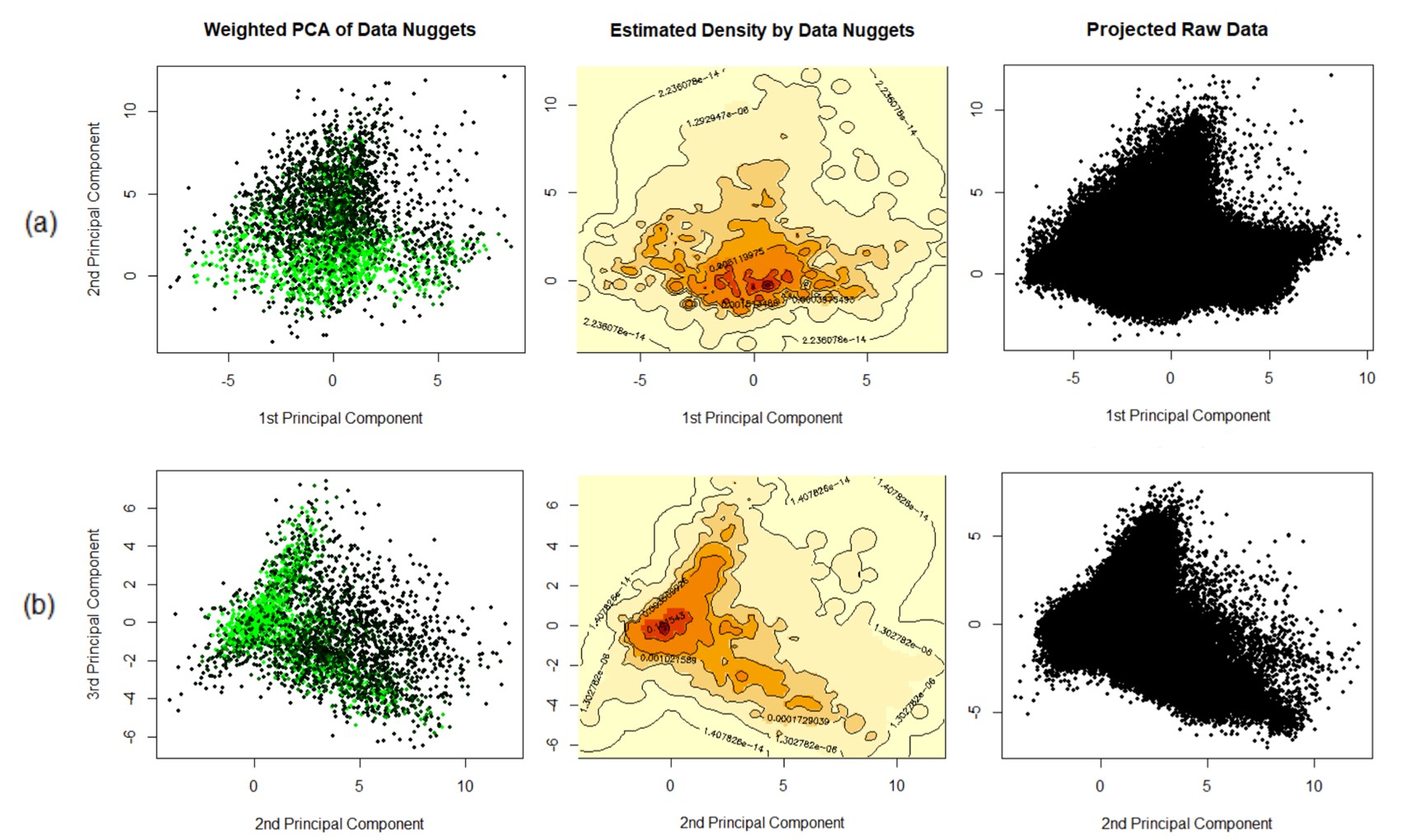}
		\caption{\small \it Weighted PCA for the data nuggets, estimated densities and the corresponding projected raw data. (a): The 1st PC vs the 2nd PC. (b): The 2nd PC vs the 3rd PC.}
\end{figure}
\begin{figure}[H]
		\centering
		\includegraphics[width=6.3in]{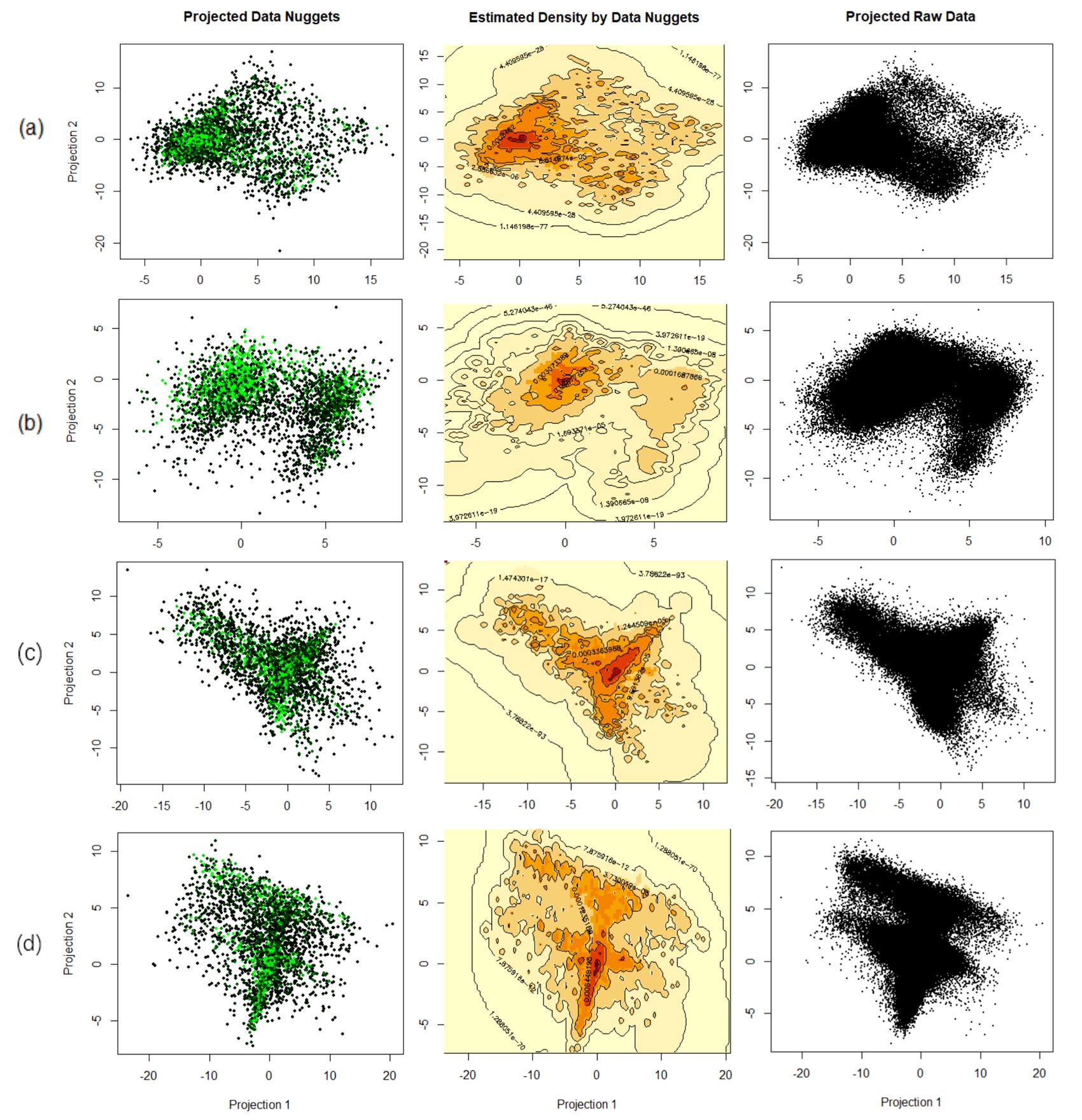}
		\caption{\small \it (a) and (b): two interesting projections with small index values. (c): a projection with a large index value. (d): the optimal projection found with the largest index value. }
\end{figure}
\begin{figure}[H]
		\centering
		\includegraphics[width=6.3in]{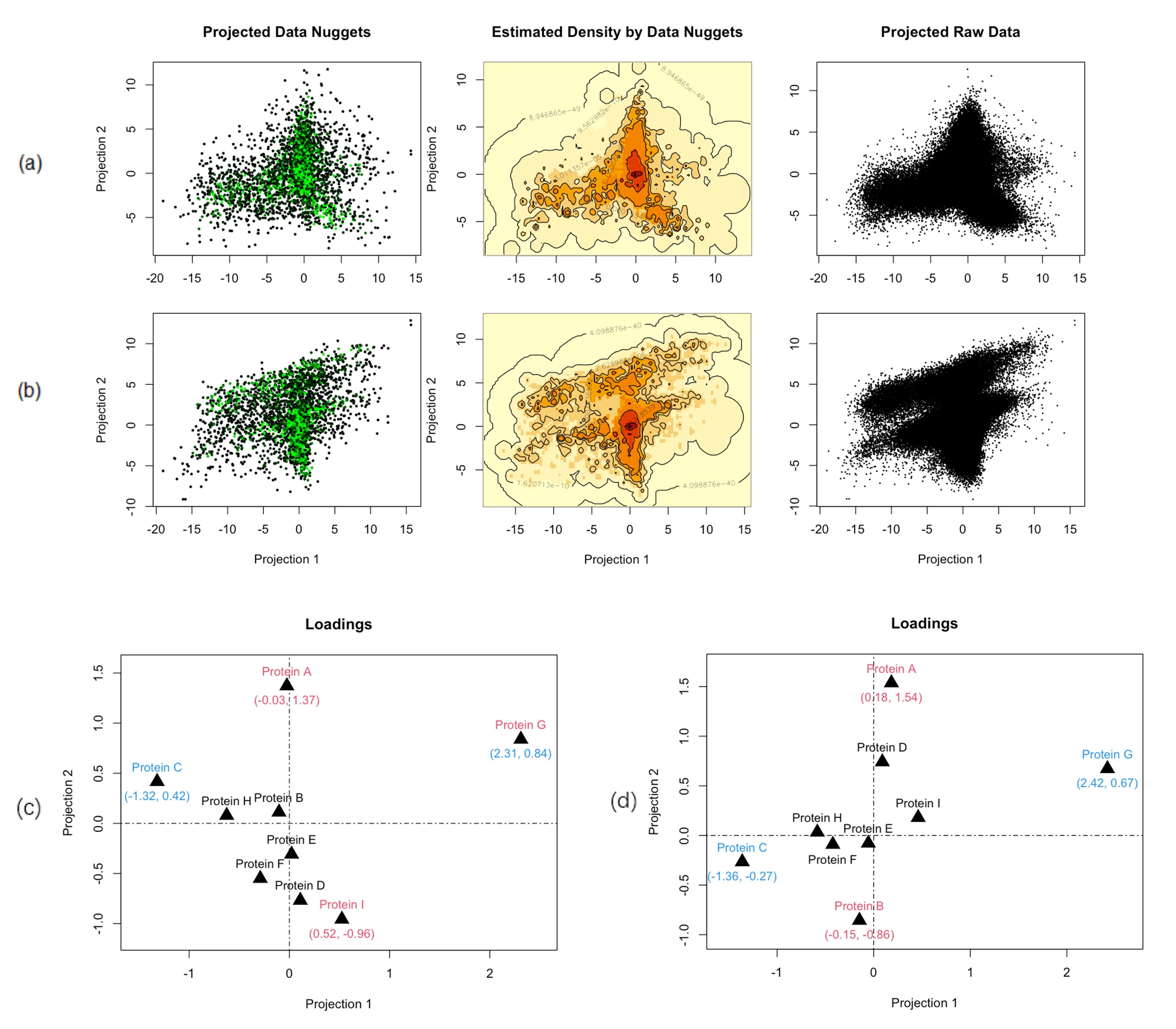}
		\caption{\small \it (a) and (b): Rotated projections by varimax rotation for optimal projections in Figure 8 (c) and (d) respectively. (c): Loadings plot for the rotated projection in (a).  (d): Loadings plot for the rotated projection in (b). }
\end{figure}

\begin{figure}[H]
		\centering
		\includegraphics[width=6.3in]{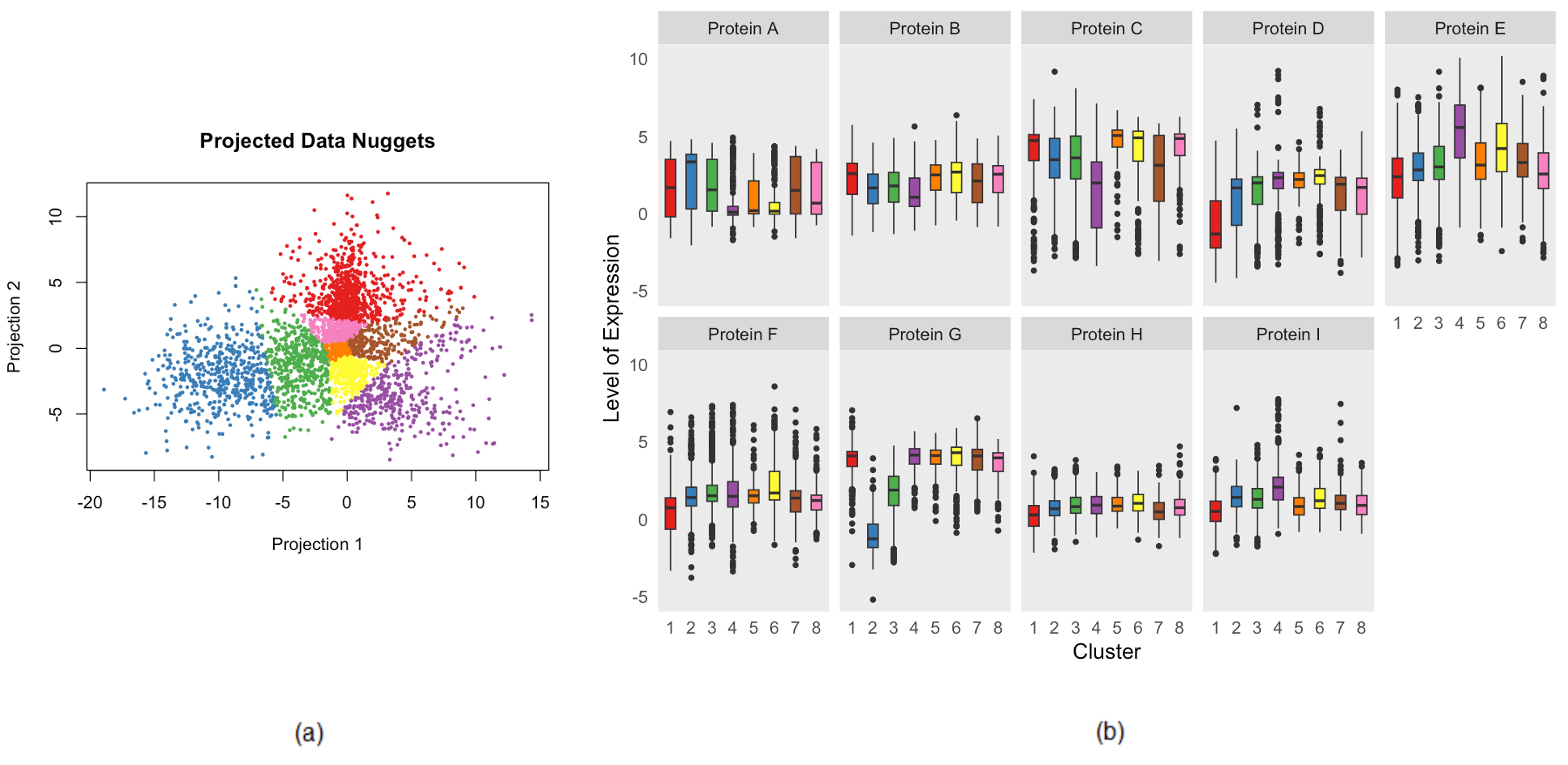}
		\caption{\small \it (a): Eight clusters by weighted k-means for the projection in Figure 8 (c).  (b): Expression levels of each protein for the eight clusters shown in (a). }
\end{figure}

\begin{figure}[H]
		\centering
		\includegraphics[width=6.3in]{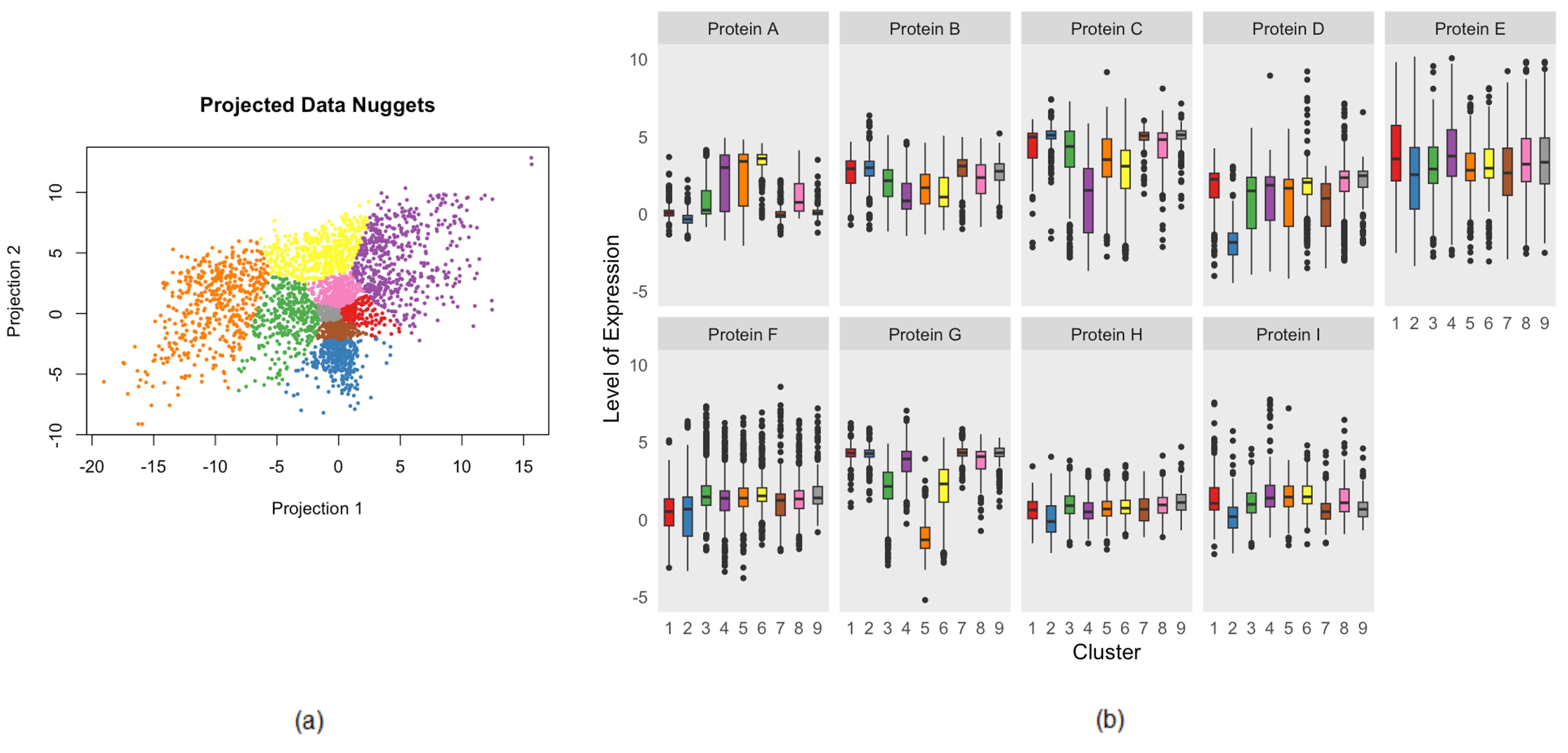}
		\caption{\small \it (a): Nine clusters by weighted k-means for the optimal projection in Figure 8 (d).  (b): Expression levels of each protein for the nine clusters shown in (a). }
\end{figure}

\section{$\!\!\!\!\!\!\!$. Discussion}  

In this paper, a new Projection Pursuit index is proposed to find hidden structures such as clusters, outliers, and other non-linear structures in extremely large datasets. It is based on a data compression method called “data-nuggets” that reduces a large dataset into a smaller collection of data nuggets that maintain the overall data structure. The new PP index is computable for big data, and simulation studies show that it is comparable to the Nature Hermite index on the full data but with much less computational cost. Optimization of the new PP index could help find the static most unusual low dimensional projection which indicates the hidden structures of big data. An application on a real large dataset from a cell flow cytometry experiment shows good performance of the new PP index. The proposed methodology helps reveal potential hidden structures in B-cell protein profiles,  which could help scientists investigate the surface protein functions of B-cells.
 
In the future, several new methods could be developed based on the proposed methodology. Firstly, the optimization algorithm for the proposed new PP index will be improved and designed to be more efficient, to find the most unusual static projection for a large data set in a short time. Secondly, the guided tour for big data will be developed based on the proposed PP index with the improved optimization method. It would help provide a new graphical tool to generate interactive, dynamic, and efficient visualization for big data and detect non-linear structures. Thirdly, other PP indices based on the data nuggets will be proposed with visualization tools, including the indices for specific structures such as clusters, holes, etc. Also, there are other potential applications of the new PP index including regarding it as a dissimilarity measure for clustering large functional data. Moreover, a differential PP index could be developed to detect changes in distributions or clusters of data between multiple groups. The index could find the optimal projection indicating the largest difference of projected data between groups. For example, for the cell flow cytometry experiments, a differential PP index could be used to find the cells that have changed their states during treatments, by detecting the changes of clusters in datasets from treatment and control groups. 

In a word, a new framework of Projection Pursuit for big data was proposed based on a data nugget method. It could help find low-dimensional projections that indicate hidden non-linear structures for large high-dimensional data with low computational cost, and is applicable to investigating large datasets in different fields.



\section*{References}
\begin{description} \itemsep=-\parsep \itemindent=-1.3cm
\item[ ] Asimov, D. (1985). The grand tour: a tool for viewing multidimensional data. \emph{SIAM journal on scientific and statistical computing}, 6(1), 128-143.

\item[ ] Buja, A., $\&$ Asimov, D. (1986). Grand tour methods: an outline. \emph{Computing Science and Statistics}, 17, 63-67.

\item[ ] Beavers, T., Cabrera, J., Lubomirski M. (2019). Data Nuggets: A Method for Reducing Big Data While Preserving Data Structure. \emph{Submitted for Publication}..

\item[ ] Beavers, T., Duan, Y., Cabrera, J., Cheng, G., Qi, K., $\&$ Lubomirski, M. (2023) datanugget: Create, and Refine Data Nuggets. R package version 1.2.4. 

\item[ ] Cherasia, K. E., Cabrera, J., Fernholz, L. T., $\&$ Fernholz, R. (2022). Data Nuggets in Supervised Learning. In \emph{Robust and Multivariate Statistical Methods: Festschrift in Honor of David E. Tyler} (pp. 429-449). Cham: Springer International Publishing.

\item[ ] Cook, D. (2009). Incorporating exploratory methods using dynamic graphics into multivariate statistics classes: curriculum development. \emph{In Quality Research in Literacy and Science Education} (pp. 337-355). Springer, Dordrecht.

\item[ ] Cook, D., Buja, A., $\&$ Cabrera, J. (1993). Projection pursuit indices based on orthonormal function expansions. \emph{Journal of Computational and Graphical Statistics}, 2(3), 225-250.

\item[ ] Cook, D., Buja, A., Cabrera, J., $\&$ Hurley, C. (1995). Grand tour and projection pursuit. \emph{Journal of Computational and Graphical Statistics}, 4(3), 155-172.

\item[ ] Cook, D., Buja, A., Lee, E. K., $\&$ Wickham, H. (2008). Grand tours, projection pursuit guided tours, and manual controls. \emph{In Handbook of data visualization} (pp. 295-314). Springer, Berlin, Heidelberg.

\item[ ] Duan, Y., Cabrera, J., Cheng, G. (2023). WCluster: Clustering and PCA with Weights, and Data Nuggets Clustering. R package version 1.2.0. 

\item[ ] Friedman, J. H. (1987). Exploratory projection pursuit. \emph{Journal of the American statistical association}, 82(397), 249-266.

\item[ ] Friedman, J. H., $\&$ Tukey, J. W. (1974). A projection pursuit algorithm for exploratory data analysis. \emph{IEEE Transactions on computers}, 100(9), 881-890.

\item[ ] Huber, P. J. (1985). Projection pursuit. \emph{The annals of Statistics}, 435-475.

\item[ ] Lee, E. K., Cook, D., Klinke, S., $\&$ Lumley, T. (2005). Projection pursuit for exploratory supervised classification. \emph{Journal of Computational and graphical Statistics}, 14(4), 831-846.

\item[ ] Swayne, D. F., Lang, D. T., Buja, A., $\&$ Cook, D. (2003). GGobi: evolving from XGobi into an extensible framework for interactive data visualization. \emph{Computational Statistics  $\&$ Data Analysis}, 43(4), 423-444.

\item[ ] Tryputsen, V., Ananth, C. V., Sargsyan, D., Kostis, J. B., Kostis, W. J., Cabrera, J. (2021). Data Nuggets for Matching Large Clinical Datasets: Application in a 350,000 Record Perinatal Database. \emph{Circulation}, 144(Suppl$\_$1) , A13403-A13403.

\item[ ] Wickham, H., Cook, D., Hofmann, H., $\&$ Buja, A. (2011). tourr: An R package for exploring multivariate data with projections. \emph{Journal of Statistical Software}, 40(2), 1-18.
 
\end{description}

\end{document}